\documentclass[aps,
 prx,amsmath,amssymb,showpacs,showkeys,longbibliography,
 reprint,
]{revtex4-1}

\usepackage{graphicx}
\usepackage{dcolumn}
\usepackage{bm}
\usepackage{color}
\usepackage[svgnames,dvipsnames]{xcolor}   
\usepackage{ulem}
\usepackage{oplotsymbl}
\usepackage{upgreek}
\usepackage{chngcntr}

\begin{document}


\title[Sample title]{A primary quantum current standard based on the Josephson and the quantum Hall effects}


\author{Sophie Djordjevic$^{1}$, Ralf Behr$^{2}$, and Wilfrid Poirier$^{1}$}
\email{wilfrid.poirier@lne.fr}
\address{$^{1}$ Laboratoire national de m\'etrologie et d'essais, 29 avenue Roger Hennequin, 78197 Trappes, France}
\address{$^{2}$ Physikalisch-Technische Bundesanstalt (PTB), Bundesallee 100, 38116 Braunschweig, Germany}

\date{\today}

\renewcommand{\figurename}{\textbf{Fig.}}
\renewcommand{\thefigure}{\arabic{figure}}
\renewcommand{\tablename}{\textbf{Table}}
\renewcommand{\thetable}{\arabic{table}}
\begin{abstract}
\normalsize
The new definition of the ampere calls for a quantum current standard able to deliver a flow of elementary charges, $e$, controlled with a relative uncertainty of $10^{-8}$. Despite many efforts, nanodevices handling electrons one by one have never demonstrated such an accuracy for a net flow. The alternative route based on applying Ohm’s law to the Josephson voltage and quantum Hall standards recently reached the target uncertainty in the milliampere range, but this was at the expense of the application of error corrections. Here, we present a new programmable quantum current generator, which combines both quantum standards and a superconducting cryogenic amplifier in a quantum electrical circuit enabling the current scaling without errors. Thanks to a full quantum instrumentation, we demonstrate the accuracy of the generated currents, in the microampere range, at quantized values, $\pm(n/p)ef_\mathrm{J}$, with relative uncertainties less than $10^{-8}$, where $n$ and $p$ are integer control parameters and $f_\mathrm{J}$ is the Josephson frequency. This experiment sets the basis of a universal quantum realization of the electrical units, for example able of improving high-value resistance measurements and bridging the gap with other quantum current sources.
\end{abstract}
\maketitle
\normalsize
Since the last revision of the International System of Units (SI) in May 20, 2019, founded on seven fixed constants of nature\cite{BrochureSI,Poirier2019}, any source generating an electric current which can be expressed in terms of $ef$, with $e$ the elementary charge and $f$ a frequency in Hz ($\mathrm{s^{-1}}$), provides a realisation of the ampere. 

Single-electron current sources (SECS) \cite{Pothier1992, Pekola2013,Scherer2019}, which are mesoscopic devices \cite{Keller1999, Pekola2008, Camarota2012, Giblin2012, Stein2015, Stein2017, Yamahata2016, Zhao2017, Bae2020}, able to handle electrons one by one at a rate $f_e$, are often presented as the most obvious way to realize the definition. However, achieving currents above 100 pA using GaAs and Si-based tunable-barrier SECS accurate to within a relative uncertainty better than $10^{-7}$ remains a very challenging goal because of increasing error rates at high frequencies ($\sim$ 1 GHz) \cite{Kataoka2011,Ahn2017}. Very recently, as a consequence of the phase-charge quantum mechanical duality in Josephson junctions (JJ), dual Shapiro steps have been evidenced in superconducting nanowires and small JJ placed in high impedance environments under microwave radiation \cite{Shaikhaidarov2022,Crescini2023,Kaap2024}. Here, the enhanced phase variance allows photon-assisted tunneling of fluxons $\upphi_0=h/2e$ ($h$ is the Planck constant) and a synchronized transfer of Cooper pairs. Sharp current steps appearing at integer multiples of $2ef_e$ in the DC current-voltage characteristics could be promising candidates as quantum sources in the nA range, although their flatness is still in debate \cite{Kurilovich2024}. More generally, for all mesoscopic current sources, the control of charge fluctuations, which are dependent on the device coupling with the electromagnetic environment, remains a crucial issue. 
\begin{figure*}[ht]
\includegraphics[width=17.5cm]{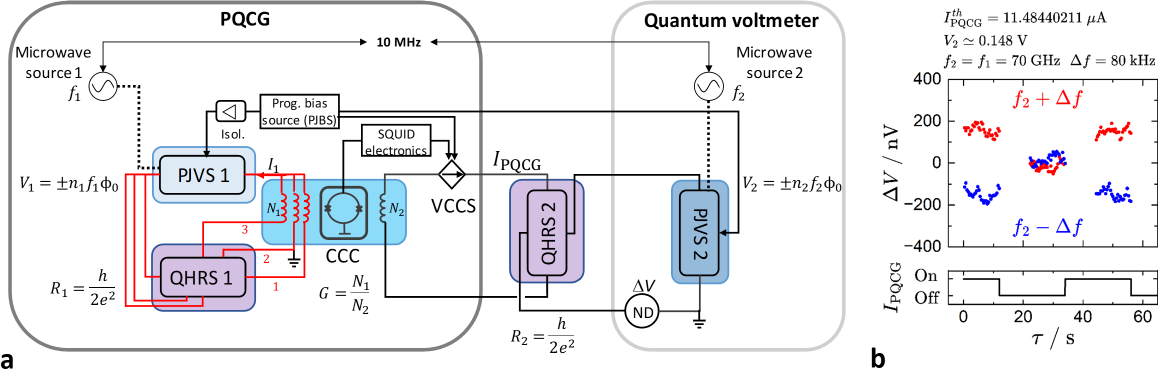}\hfill
\caption{\textbf{Experimental set-up for the accuracy test of the PQCG.} \textbf{a} The PQCG is composed of $\mathrm{PJVS_1}$ and $\mathrm{QHRS_1}$ connected through a triple connection (red lines 1, 2, 3) ensured via the windings of $N_1$ turns of a new CCC, which are used to measure and amplify the quantized current $I_1$. The shields at ground potential of the high and low potential cables are not shown on the scheme. The SQUID detector of the CCC feedbacks on an external current source (VCCS) which maintains the current $I_\mathrm{PQCG}$ into the series connection of $\mathrm{QHRS_2}$ and of the winding of $N_2$ turns. To reduce the finite gain error of the SQUID, a voltage source of the programmable Josephson bias source (PJBS) drives the VCCS and preadjusts the output current, such that the SQUID feedbacks only on a small fraction of $I_\mathrm{PQCG}$ (see Uncertainties section in Methods). The voltage drop at $\mathrm{QHRS_2}$ is measured in a quantum voltmeter configuration made of $\mathrm{PJVS_2}$ and a null detector, ND (EM Electronics Model N11). The frequency of the Josephson microwave signals are referenced to a 10~MHz frequency standard linked to an atomic clock time base. Both PJVS are controlled and synchronized using the same PJBS. For measurements performed with $n_1=4096$, a decoupling stage ensures the electrical isolation between the two PJVS so that the PQCG and the quantum voltmeter can each be grounded. For $n_1=1920$, no decoupling stage is used implying that only $\mathrm{PJVS_1}$ is grounded. Four different cryostats are used to cool down quantum devices (represented by purple, blue gray, cyan and light blue colors). \textbf{b} ND recordings, $\Delta V$, for frequencies $f_2+\Delta f$ (red circle) and $f_2-\Delta f$ (blue circle) as a function of time $\tau$ while the current $I_\mathrm{PQCG}$ is periodically switching on and off (I$_{+}$ measurement protocol).}\label{fig.1}
\end{figure*}

Concurrently, another route to the SI realization consists in applying Ohm's law to the Josephson voltage and quantum Hall resistance standards, since the Josephson effect \cite{Josephson1962} and the quantum Hall effect \cite{Klitzing80} now provide direct and universal realizations of the volt and the ohm from $h/2e$ and $h/e^2$ constants, respectively \cite{Poirier2019}, with a $10^{-9}$ measurement uncertainty. The high accuracy of the Josephson voltage standards, which are series arrays of JJ, relies on the phase rigidity of macroscopic superconductors. Under application of dc current bias and a microwave radiation $f_\mathrm{J}$, the transfer through each JJ of one fluxon per period of the microwave tone is ensured and results in a quantized voltage $V=n_\mathrm{J}\upphi_0 f_\mathrm{J}$ \cite{Shapiro63}, where $n_\mathrm{J}$ is the number of JJ. If this quantized voltage can be accurately applied to a quantum Hall resistance standard (QHRS) in the $\nu$=2 Landau level filling factor, taking advantage of the charge rigidity of the quantum Hall edge states, $n_\mathrm{J}e$ charges are transferred at a rate $f_\mathrm{J}$ through the QHRS of $h/2e^{2}$ resistance. Hence, a current $n_\mathrm{J}ef_\mathrm{J}$, easily reaching microamperes, can be generated. Recently, a calculable current of 1 $\upmu$A generated from the series connection of both quantum standards has been measured with a relative uncertainty of $1.3\times10^{-7}$ \cite{Chae2022,Kaneko2024}, but the accuracy was reached owing to the very low-resistance of the ammeter ($\sim0.1$ m$\Omega$). The main issue is therefore to implement the accurate series connection of the two quantum standards while realizing a true current source. The programmable quantum current generator (PQCG)~\cite{Poirier2014,Brun-Picard2016,Djordjevic2021} has addressed this issue by locking an electron flow to the current circulating in the loop formed by the quantum standards, with the help of a superconducting amplifier, allowing simultaneously the scaling over a wider range of current values. Its accuracy was demonstrated in the milliampere range with a relative measurement uncertainty of $10^{-8}$. This result was however obtained at the expense of corrections of the order of a few parts in $10^{7}$, originating from the non-ideal connection of the quantum standards, and determined after time consuming additional measurements. This impaired the final uncertainty, the simplicity and hence the full potential of the new quantum current standard.

Here, we report on a next-generation PQCG able of delivering lower-noise currents, at the theoretical quantized values driven by the Josephson frequency, without any classical correction. This performance results from the implementation of a three-terminal connection of the QHRS, which allows a highly-accurate application of the Josephson voltage to the quantized resistance. We demonstrate the realization of the ampere with relative uncertainties below $10^{-8}$ for different current levels, filling the gap between the milliampere range and the microampere range.This is achieved using a full quantum instrumentation made of five quantum devices, which opens the way to the realization of several electrical units in a single experiment.\\ \\
\Large\textbf{Results}\\ \\
\normalsize
\textbf{Next-generation PQCG}\\
Fig.\ref{fig.1}a shows the implementation (see also Quantum devices section in Methods) of two programmable Josephson voltage standards (PJVS), two QHRS and a cryogenic current comparator (CCC). The two PJVS are binary divided 1 V $\mathrm{Nb/Nb_xSi_{1-x}/Nb}$ series arrays \cite{Behr2012}, both having a total of 8192 JJ and working around 70 GHz. The voltage of the two PJVS are given by $\pm n_{1,2}\upphi_0 f_\mathrm{1,2}$ with $n_{1,2}$ the number of JJ biased on the $\pm$1 Shapiro steps. The two QHRS are both GaAs/AlGaAs heterostructures \cite{Piquemal1993} of quantized resistance $R_1=R_2=h/2e^{2}$. The CCC is a dc current transformer\cite{Harvey1972}, made of several superconducting windings of different number of turns, able to compare currents with a great accuracy (below one part in $10^9$) and sensitivity (80 pA$\cdot$turns/$\mathrm{Hz^{1/2}}$) owing to Amp\`ere's law and Meissner effect. 

The new version of the PQCG is composed of $\mathrm{PJVS_1}$ \cite{Behr2012} connected to $\mathrm{QHRS_1}$ with a triple connection (see Quantum devices section in Methods) ensured through three identical windings of $N_1$ turns of a new specially designed CCC (see CCC section in Methods). This connection technique \cite{Delahaye1993} reduces the impact of the series resistances to an insignificant effect. More precisely, one current contact and two voltage contacts of the same equipotential of $\mathrm{QHRS_1}$ are connected all together at each superconducting pads of $\mathrm{PJVS_1}$. Because of the topological properties of Hall edge-states, namely their chirality, their $h/e^2$ two-wire resistance and their immunity against backscattering \cite{Buttiker1988}, the current flowing through the third contact is only a fraction $(r/R_1)^2$ of the current circulating in the first one, where $r$ is the typical resistance of the connections. The resistance seen by $\mathrm{PJVS_1}$ is close to $h/2e^2$ within a typical small correction of order $(r/R_1)^3$. It results that the total current circulating $I_1$, is close to $\pm n_\mathrm{1}ef_1$ within 1.5 parts in $10^{10}$ for series resistance values lower than 5 ohms (see Multiple series connection section in Methods). Compared to \cite{Brun-Picard2016}, where the double connection required the application of a relative correction to the current of a few $10^{-7}$, the operation of the PQCG is simplified since no correction is necessary here. The quantized current $I_1$, divided in the three connections, is measured by the three identical windings of $N_1$ turns. A DC SQUID is used to detect the unbalance ampere$\cdot$turns in the different windings of the new CCC. It feedbacks on the new battery-powered voltage controlled current source (VCCS), which supplies a winding of $N_2$ turns with current $I_\mathrm{PQCG}$ in order to maintain the ampere$\cdot$turns balance $N_\mathrm{1}I_\mathrm{1}-N_\mathrm{2}I_\mathrm{PQCG}=0$. It results that the PQCG is able to output a current equal to:
\begin{equation}
I_\mathrm{PQCG}^\mathrm{th}=\pm(N_\mathrm{1}/N_\mathrm{2})n_\mathrm{1}ef_\mathrm{1},
\end{equation}\label{IPQCG}
to within an estimated Type B relative uncertainty of 2 parts in $10^9$ (see Table \ref{Type-B PQCG} in Methods). In practice, the CCC gain $G=N_1/N_2$ can span two orders of magnitude on either side of the unity, allowing the generation of currents with values from nanoampere to milliampere.\\ \\
\textbf{Accuracy test principle}\\
No matter how accurately the Type B uncertainty may be estimated, a metrological requirement is to check that the experimental current $I_\mathrm{PQCG}$ is given by the relationship (1) using another quantum measurement method through an accuracy test. The output current is determined by feeding $\mathrm{QHRS_2}$, and by measuring the voltage drop, $V_2$, at its Hall terminals using a quantum voltmeter (Fig.1 and Supplementary Fig.1). The latter is made of $\mathrm{PJVS_2}$ and an analog null detector (ND), which measures the voltage difference $\Delta V$. From Kirchhoff's voltage law, $I_\mathrm{PQCG}$ is determined according to the expression :
\begin{equation}
V_2=R_\mathrm{2}I_\mathrm{PQCG}=\pm\upphi_0n_\mathrm{2}f_\mathrm{2}-\Delta V, 
\end{equation}
with $\Delta V=0$ ideally at the equilibrium frequency $f_\mathrm{2}^{eq}$. Using a quantized resistance $R_\mathrm{2}=h/2e^2$ ($\sim 12.9~\mathrm{k}\Omega$), about 129 times higher than in \cite{Brun-Picard2016}, allows increasing the signal-to-noise ratio while eliminating an extra resistance calibration. 
The relative deviation of the measured current to the theoretical one, $\Delta I/I=(I_\mathrm{PQCG}/I_\mathrm{PQCG}^\mathrm{th})-1$, is given by $\Delta I/I=\frac{n_\mathrm{2}f_\mathrm{2}^{eq}}{Gn_\mathrm{1}f_\mathrm{1}}-1$. In practice, the CCC gain and the numbers of JJ are chosen so that $Gn_\mathrm{1}=n_\mathrm{2}$. Thus, the nominal relative deviation can be expressed as:
\begin{equation}
\Delta I/I=\frac{f_\mathrm{2}^\mathrm{eq}}{f_\mathrm{1}}-1.
\end{equation}
This relationship (3) clearly emphasizes that the accuracy test resumes to the determination of the equilibrium frequency $f_\mathrm{2}^\mathrm{eq}$ which should be equal to $f_1$, if the current is accurately quantized, but can correspond to another value in case of a systematic error. However, noise and offset drifts can make it difficult to find the frequency corresponding precisely to $\Delta V=0$. Here, instead of performing a single measurement at frequency $f_2$ close to $f_\mathrm{2}^\mathrm{eq}\sim f_1$, and measuring a voltage signal hidden by the noise, two successive voltage mean values, $\overline{\Delta V_{f_2^{+}}}$ and $\overline{\Delta V_{f_2^{-}}}$, are measured at two detuned frequencies $f_\mathrm{2}^+=f_\mathrm{2}+\Delta f$ and $f_\mathrm{2}^-=f_\mathrm{2}-\Delta f$, respectively, where $\Delta f$ is set to 40 kHz or 80 kHz in our experiments (Fig.\ref{fig.1}b and Supplementary Fig.2a). The measurement of significant voltage signals is a straightforward way to evaluate the signal to noise ratio and quickly detect any anomalous event such as trapped magnetic flux either in the SQUID or in the JJ. In order to mitigate the effect of offsets, drifts and 1/$f$ noise, each voltage mean value is obtained from a measurement series consisting in periodically either switching on and off the current with $I_\mathrm{PQCG}>0$ (I$_+$) or $<0$ (I$_-$), or completely reversing the current (I$_\pm$). The equilibrium frequency is then determined from :
\begin{equation}
f_\mathrm{2}^\mathrm{eq}=\frac{f_\mathrm{2}^-\overline{\Delta V_{f_2^{+}}}-f_\mathrm{2}^+\overline{\Delta V_{f_2^{-}}}}{(\overline{\Delta V_{f_2^{+}}}-\overline{\Delta V_{f_2^{-}}})}, 
\end{equation}
which implies voltage ratios only. The two successive voltage measurements allow an in-situ and short-time calibration of the nanovoltmeter gain from the Josephson frequency, overcoming its lack of stability over time. Finally, the determination of $\Delta I/I$ does not require any calibration, resulting in a reduced Type B standard uncertainty, $u^\mathrm{B}$, of $2.1\times10^{-9}$ (see Table \ref{Type-B PQCG} in Methods). The Type A standard uncertainty of $\Delta I/I$, $u^\mathrm{A}=\frac{u^\mathrm{A}(f_\mathrm{2}^\mathrm{eq})}{f_\mathrm{1}}$ is determined from the standard deviations of the mean of the voltage series $\Delta V_{f_2^{+}}$ and $\Delta V_{f_2^{-}}$ (see Uncertainties section in Methods). This is justified by the calculation of the relative Allan deviation~\cite{Brun-Picard2016} showing a dominant white noise contribution (Supplementary Fig.2b). 
\begin{figure*}[ht]
\includegraphics[width=17.5cm]{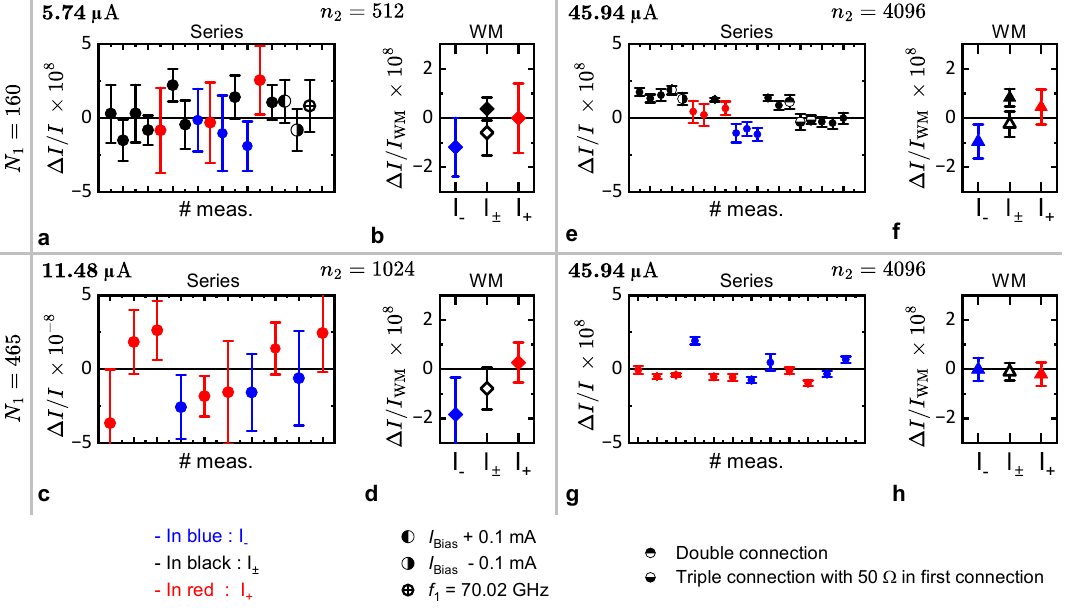}\hfill
\caption{\textbf{Accuracy tests of the PQCG.} \textbf{a, c, e, g} Series of relative deviations $\Delta I/I$ with $n_1=4096$ measured at 5.742201056 $\upmu$A using $N_1=160$ (\textbf{a}), at 11.48440211 $\upmu$A using $N_1=465$ (\textbf{c}), at 45.93760845 $\upmu$A using $N_1=160$ (\textbf{e}) and at 45.93760845 $\upmu$A using $N_1=465$ (\textbf{g}). Measurements were carried out over one or two days, using I$_+$, I$_-$ or I$_\pm$ measurement protocols (circle). Some measurements are performed with $I_\mathrm{bias}$ shifted by +0.1 mA and -0.1 mA (\textbf{a} and \textbf{e}), with a frequency $f_\mathrm{1}=70.02$ GHz (\textbf{a}), using the double connection (applying the corresponding $\alpha_2=(2.344\pm 0.037)\times10^{-7}$ correction) and using the triple connection with 50 $\Omega$ inserted into the first connection (applying the corresponding $\alpha_3=(1.4\pm0.02)\times10^{-9}$ correction). All uncertainties correspond to one standard deviation, i.e., are given with a coverage factor $k$=1. Error bars correspond to type A standard uncertainties, $u^\mathrm{A}$. \textbf{b, d, f, h} Weighted mean values $\Delta I/I_\mathrm{WM}$ of the series measured at 5.74 $\upmu$A (\textbf{b}), at 11.48 $\upmu$A (\textbf{d}), at 45.94 $\upmu$A using $N_1=160$ (\textbf{f}) and at 45.94 $\upmu$A using $N_1=465$ (\textbf{h}). In \textbf{b} and \textbf{d}, weighted mean values for I$_+$, I$_-$, I$_\pm$ (filled diamond) are calculated from the $u^\mathrm{A}$, mean value of I$_+$ and I$_-$ results (open diamond). In \textbf{f} and \textbf{h}, weighted mean values for I$_+$, I$_-$, I$_\pm$ (filled triangle) are calculated from $\sqrt{(u^\mathrm{A})^2+(u^\mathrm{A}_\mathrm{it-noise})^2}$, where $u^\mathrm{A}_\mathrm{it-noise}$ accounts for the observed residual intra-day noise, mean value of I$_+$ and I$_-$ results (open triangle). Error bars correspond to combined standard uncertainties $u^\mathrm{c}_\mathrm{WM}$ (see Uncertainties section in Methods). Other parameter: $N_2$ is fixed such as $G=N_1/N_2=n_2/n_1$.}\label{fig.2}
\end{figure*} \\ \\
\textbf{Quantized current accuracy}\\
Measurements of $\Delta I/I$ were performed at four different values of current 5.74~$\upmu$A, 11.48~$\upmu$A, 45.94~$\upmu$A and 57.42~$\upmu$A, using a primary current $I_\mathrm{1}$ of 45.94~$\upmu$A obtained with $n_\mathrm{1}=4096$. This large current improves the operational margins of the PQCG compared to~\cite{Brun-Picard2016} and increases the signal-to-noise ratio while ensuring a perfect quantization of the Hall resistance of the $\mathrm{QHRS_1}$ device. Each measurement series was typically carried out over one day using I$_+$, I$_-$, and I$_\pm$ measurement protocols to reveal any systematic effect related to the current direction. Note that implementing complete current reversals I$_\pm$ required the reduction of the noise in the circuit~\cite{Djordjevic2021} (see CCC section in Methods). 
Measurements were performed with $N_1=160$ or also with $465$ to test the effect of the number of ampere$\cdot$turn. However, the downside of the latter configuration is the higher instability of the feedback loop encountered during the current reversals which prevented the use of the I$_{\pm}$ measurement protocol. The different output currents were obtained by changing $N_2$ from 80 to 1860. Other PQCG parameters are reported in Table \ref{SettingAccuracyTest} in Methods.

At the lower current values of 5.74 $\upmu$A (Fig.\ref{fig.2}a) and 11.48 $\upmu$A (Fig.\ref{fig.2}c), discrepancies of $\Delta I/I$ are covered by Type A uncertainties, $u^\mathrm{A}$, ranging from 1 to $3\times10^{-8}$. Weighted means, $\Delta I/I_\mathrm{WM}$, for each measurement protocol I$_+$, I$_-$ and I$_\pm$, reported in Fig.\ref{fig.2}b and Fig.\ref{fig.2}d, show that there is no significant deviation of the current from its theoretical value within combined measurement uncertainties of about $10^{-8}$. Besides, the mean value of $\Delta I/I_\mathrm{WM}(\mathrm{I}_+)$ and $\Delta I/I_\mathrm{WM}(\mathrm{I}_-)$ is clearly in agreement with $\Delta I/I_\mathrm{WM}(\mathrm{I}_\pm)$ at 5.74 $\upmu$A, which confirms the equivalence of averaging measurements carried out using the I$_+$ and I$_-$ protocols with the measurement obtained using the I$_\pm$ protocol. Combining the different results (see Weighted mean values section in Methods), one obtains the relative deviations, for the equivalent protocol I$_\pm$, $\overline{\Delta I/I}=(2\pm4.3)\times10^{-9}$ and $\overline{\Delta I/I}=(-7.9\pm 8.6)\times10^{-9}$ at current levels of 5.74 $\upmu$A and 11.48 $\upmu$A, respectively.

At the higher current value of 45.94 $\upmu$A, $\Delta I/I$ measurements, reported in Fig.\ref{fig.2}e and Fig.\ref{fig.2}g, are characterized by smaller Type A standard uncertainties, $u^\mathrm{A}$, in agreement with the larger voltage drop at the terminals of $\mathrm{QHRS_2}$ (about 0.59 V). As expected, even smaller uncertainties are observed for $N_\mathrm{1}=465$ than for $N_\mathrm{1}=160$ due to an enhanced ampere.turns value $N_1I_1$. The lower uncertainties permit the scrutiny of small but significant deviations revealing intra-day noise at the $10^{-8}$ level, the origin of which has not been clearly identified yet. To account for this, the standard uncertainty of each $\Delta I/I$ measurement is increased of an additional Type A uncertainty component $u^\mathrm{A}_\mathrm{id-noise}=10^{-8}$ chosen so that the $\chi^2$ criterion is fulfilled (see Uncertainties section in Methods). The resulting weighted mean values, $\Delta I/I_\mathrm{WM}$, shown in Fig.\ref{fig.2}f and Fig.\ref{fig.2}h, do not reveal any significant deviation from zero with regards to the combined measurement uncertainties of only a few $10^{-9}$. Combining results obtained using the different measurement protocols, one obtains $\overline{\Delta I/I}=(5.4\pm3.1)\times10^{-9}$ for $N_\mathrm{1}=160$ and $\overline{\Delta I/I}=(-1.1\pm3.6)\times10^{-9}$ for $N_\mathrm{1}=465$. Hence, on average over a day, the current delivered by the PQCG is quantized and the deviation from zero is covered by an uncertainty of about $4\times10^{-9}$. At shorter terms, the uncertainty due to the intra-day noise does not average out and the combined uncertainty is $\simeq10^{-8}$. Generally, one might guess a small discrepancy between measurements performed using either I$_+$ or I$_-$ protocols, in Fig.\ref{fig.2}b, c and d, which could come from a small Peltier type effect. However using the protocol I$_\pm$ cancels this potential effect.  
\begin{figure}[h]
\includegraphics[width=8.5cm]{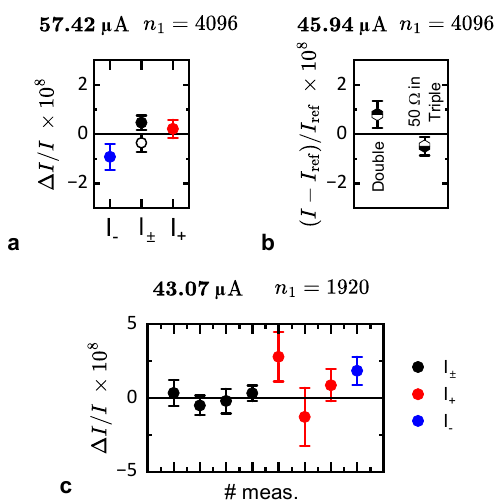}\hfill
\caption{\textbf{Robustness tests of the PQCG accuracy}. \textbf{a} Relative deviations $\Delta I/I$ at 57.42201056 $\upmu$A measured with $n_1=4096$ for I$_+$, I$_-$ and I$_\pm$ protocols, mean value of I$_+$ and I$_-$ results (open circle). Errors bars corresponds to combined uncertainties (see Uncertainties section in Methods). \textbf{b} Relative current shift from a reference value $I_\mathrm{ref}$, in back and forth measurements testing the effectiveness of the multiple connection (data extracted from Fig.\ref{fig.2}e (see Combined results section in Methods) at 45.94 $\upmu$A): measurements using the double connection with the $2^\mathrm{nd}$ order correction ($\alpha_2$) applied, using the triple connection with 50 $\Omega$ inserted in the first connection with third order correction $\alpha_3$ applied. The results are consistent within the Type A standard uncertainty. This confirms the accuracy of the cable corrections when applied and the efficiency of the triple connection against large cable resistance value. \textbf{c} Series of $\Delta I/I$ measurements carried out at 43.06650792 $\upmu$A for I$_\pm$, I$_+$ and I$_-$ measurement protocols using a second $\mathrm{PJVS_1}$ configuration with $n_1=1920$ and $G=N_1/N_2=160/80$. Errors bars corresponds to $u^\mathrm{A}$ uncertainties. Discrepancies are within less than two parts in $10^8$. This demonstrates that the PQCG can be accurate within this uncertainty even though the biasing scheme is less robust with respect to trapping magnetic flux (see Quantum devices section in Methods).}\label{DataCompl1et2}
\end{figure}\\ \\
Similar results are obtained for $\overline{\Delta I/I}$ at current value of 57.42~$\upmu$A ($N_1=160$), giving $(1.9\pm2.6)\times10^{-9}$ (Fig.\ref{DataCompl1et2}a). The margins over which the current values remain quantized at the same level of uncertainty have been tested in different situations. No significant deviation of the generated current is measured when shifting by $\pm0.1$~mA the Josephson bias current of PJVS$_1$, $I_\mathrm{bias}$ (shown in Fig.\ref{fig.2}a at 5.74~$\upmu$A and in Fig.\ref{fig.2}e at 45.94~$\upmu$A) or by varying the PJVS$_1$ frequency from 70~GHz to 70.02~GHz (Fig.\ref{fig.2}a). The efficiency of the triple connection against large cable resistance value and the accuracy of the cable corrections, when applied, have been demonstrated by inserting a large resistance (50~$\Omega$) into the first connection of the triple connection scheme, and by application of the cable correction in the double connection scheme (Fig.\ref{fig.2}e and Fig.\ref{DataCompl1et2}b, respectively). Finally, at 43.07~$\upmu$A, another connection scheme (see Quantum devices section in Methods), including JJ into the triple connection of $\mathrm{PJVS_1}$ with $n_1=1920$, although less reliable with respect to magnetic flux trapping, confirms an accuracy at a level of a few parts in $10^8$ (Fig.\ref{DataCompl1et2}c).
\begin{figure}[ht]
\includegraphics[width=8.5cm]{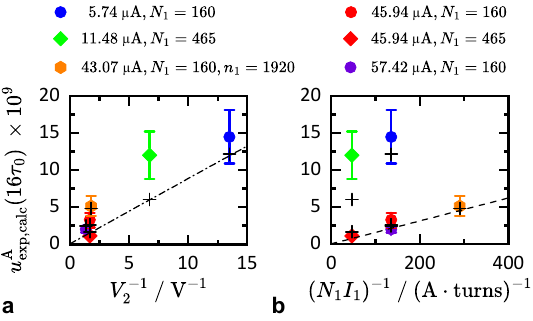}\hfill
\caption{\textbf{Noise analysis.} \textbf{a, b} Experimental uncertainties, $u^\mathrm{A}_\mathrm{exp}$, determined from all data for a normalized measurement time ($\tau=16\tau_0\sim$ 18 min) and the I$_\pm$ measurement protocol, as a function of $1/ V_2$ (\textbf{a}) and $1/N_1I_1$ (\textbf{b}). Calculated uncertainties, $u^\mathrm{A}_\mathrm{calc}$ (cross). Functions 8.8$\times 10^{-10}/V_2$ (dash-dotted line in \textbf{a}) and $1.6\times10^{-11}/(N_1I_1)$ (dash-line in \textbf{b}) represent $u^\mathrm{A}_\mathrm{calc}(\tau_m)$ calculated for $S_{\upphi}=0$ and $S_V=0$, respectively.}\label{fig.3}
\end{figure}\\ \\
\textbf{Evaluation of short-term noise sources}\\
The analysis of the $u^\mathrm{A}$ uncertainties provide a further insight into the understanding of the experiment. Fig.\ref{fig.3}a and b show averages, $u^\mathrm{A}_\mathrm{exp}(\tau_m)$, of the $u^\mathrm{A}$ uncertainties measured in the different accuracy tests, after normalization to the same measurement time $\tau_m=16\tau_0$ (with $\tau_0=66$~s, $\tau_m\sim$ 18 min) and to the same measurement protocol I$_\pm$, as a function of $1/V_2$ and $1/N_1I_1$, respectively. For comparison, they also report the theoretical standard uncertainties, $u^\mathrm{A}_\mathrm{calc}(\tau_m)=\frac{\sqrt{3}}{16}\sqrt{\frac{S_{\Delta I/I}}{\tau_A}}$, calculated using $\tau_A=12$~s from the $\Delta I/I$ noise density (see Experimental and Calculated standard uncertainties section in Methods) :
\begin{equation}
\sqrt{S_{\Delta I/I}}=\sqrt{\frac{S_V}{V_2^2}+(\frac{1}{N_1I_1}\gamma_\mathrm{CCC})^2S_{\upphi}},
\end{equation}
considering a magnetic flux noise detected by the SQUID of $\sqrt{S_{\upphi}}=62~\upmu\upphi_0/\mathrm{Hz}^{1/2}$, a voltage noise of $\sqrt{S_V}=\mathrm{28~nV/Hz^{1/2}}$ and using the CCC sensitivity $\gamma_\mathrm{CCC}=8~\upmu$A$\cdot$turn/$\mathrm{\upphi}_0$. These values give a very satisfactory agreement with the experimental data. The $\sqrt{S_{\upphi}}$ value is compatible with the noise spectrum of the CCC (see CCC section in Methods). The $\sqrt{S_V}$ value is higher than the nominal voltage noise expected from the ND (EM electronics N11) for a 13 k$\Omega$ resistor. As it was previously observed~\cite{DrungStorm2011}, the manifestation of large transient voltages at the input of the null detector during current switchings is a possible explanation for larger scattering of voltage measurements, but we cannot rule out the capture of external noise by the primary loop~\cite{Djordjevic2021}. In both cases, there is room for improvement in the future. Fig.\ref{fig.3}a confirms that, at low $V_2$, the main noise contribution comes from the voltage noise of the quantum voltmeter, as emphasized by the $1/V_2$ dependence which is reproduced by $u^\mathrm{A}_\mathrm{calc}(\tau_m)$ calculated for $S_{\upphi}=0$. On the other hand, at higher $V_2$ (Fig.\ref{fig.3}b), the number of ampere$\cdot$turns becomes the dominant parameter and the experimental uncertainties follow the $1/N_1I_1$ dependence of $u^\mathrm{A}_\mathrm{calc}(\tau_m)$ calculated for $S_{V}=0$. One can therefore deduce the Type A uncertainty contributions of the PQCG and the quantum voltmeter, which amount to $1.6\times10^{-11}/(N_1I_1)$ and 8.8$\times 10^{-10}/V_2$, respectively. The low value of $7.3\times10^{-10}$ calculated for $N_1=465$ and $I_1=45.94~\upmu$A for the PQCG itself allows considering the generation of even smaller currents. However, the demonstration of their accuracy would require the increase of $R_2$, by using a series quantum Hall array of 1.29~$\mathrm{M\Omega}$ resistance \cite{Poirier2004} to increase $V_2$.
\begin{figure}[ht]
\includegraphics[width=8.5cm]{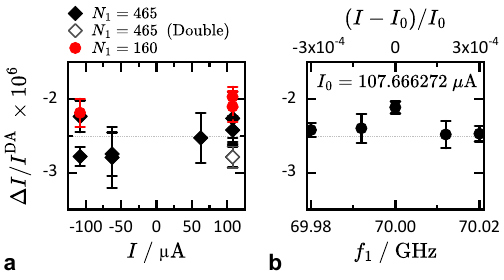}\hfill
\caption{\textbf{Calibration of a digital ammeter HP3458A}. \textbf{a} Relative deviation, $\Delta I/I^\mathrm{DA}$, as a function of current value in 100 $\upmu$A range for different PQCG configurations using $n_1=1920$: $N_1=465$, $N_2=93, 160$ (black diamond), double connection with $N_1=465$, $N_2=93$ and application the correction $\alpha_2$ (open black diamond), $N_1=160$, $N_2=32$ (red circle). Calibrations performed at $\pm$107.7 $\upmu$A using either $N_1$=465 or $N_1$=160, or using the double connection scheme and applying the cable correction $\alpha_2$ are all in agreement. \textbf{b} $\Delta I/I^\mathrm{DA}$ as a function of the Josephson frequency $f_1$, or relative deviation of current around $I_0=107.666272$ $\upmu$A (black circle).}\label{fig.4}
\end{figure} \\ \\
\textbf{Application to an ammeter calibration}\\
Fig.\ref{fig.4}a shows the relative deviations $\Delta I/I^\mathrm{DA}$ between the currents measured by the DA and the quantized currents generated by the PQCG in the 100 $\upmu$A range using the second configuration of $\mathrm{PJVS_1}$ with $n_1=1920$ (Supplementary Table.1). The coarse adjustment of the quantized current, about $\pm$107.7$~\upmu$A and $\pm$62.6$~\upmu$A, is done by using $G=465/93$ (or 160/32) and $G=465/160$, respectively. Using either I$_+$ or I$_-$ measurement protocols (Supplementary Fig.3a), Allan deviation shows that the Type A relative uncertainty for the $\tau_m=144$~s measurement time amounts to about $2\times10^{-7}$ at $\pm$107.7$~\upmu$A (Supplementary Fig.3b). Data demonstrate that the DA is reproducible over the current range within about 5 parts in $10^7$, similar to results obtained in the milliampere range~\cite{Brun-Picard2016}. Finally, Fig.\ref{fig.4}b illustrates the possibility of a fine tuning of the current by varying $f_\mathrm{J}$ from 69.98 to 70.02~GHz, which represents a relative shift of the quantized current of $\pm 3\times10^{-4}$ around $107.7$~$\upmu$A.\\ \\ \\ 
\Large{\textbf{Discussion}}\\ \\ 
\normalsize
\textbf{Quantum current standard: state-of-the-art}\\
We have demonstrated the accuracy of the flow rate of electrons generated by the new generation PQCG, at current values bridging the gap between the microampere range and the milliampere range with relative uncertainties $\leq10^{-8}$, as summarized in Fig.\ref{fig.5}. The results support our estimation of the Type B uncertainty budget. Moreover, Type A uncertainties of only a few $10^{-9}$ have been measured. These progress stem from eliminating the need of any classical correction, improving the signal-to-noise ratio, extending the operating margins and applying new measurement protocols based on tuning Josephson frequencies. These results pave the way for a quantum current standard that is as accurate as voltage and resistance standards, and similarly requires only verification of quantization criteria~\cite{Brun-Picard2016,Djordjevic2021,Delahaye2003} when used for current calibration.
\begin{figure}[ht] 
\includegraphics[width=8.5cm]{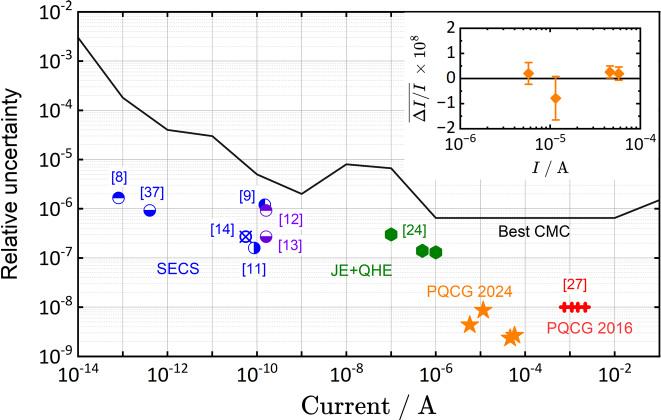}\hfill
\caption{\textbf{State-of-the-art of the accuracy of quantum current sources and current generation}. Comparison between the relative uncertainty achieved in the generation of currents traceable to the ampere using the PQCG (this work and \cite{Brun-Picard2016}) and others quantum current sources: metallic (\cite{Camarota2012},\cite{Keller2007}), GaAs (\cite{Giblin2012},\cite{Stein2017},\cite{Bae2020}) and Si (\cite{Yamahata2016},\cite{Zhao2017}) SECS, series connection of QHRS array and PJVS for generating (\cite{Chae2022}) current. CMCs (black line)\cite{BIPMCMC}. \textbf{Inset}: $\overline{\Delta I/I}$ measurements performed in this work (orange diamond) as a function of current value $I$. The value at 45.94 $\upmu$A corresponds to the weighted mean of $\overline{\Delta I/I}$ values obtained using $N_1=160$ and $N_1=465$.}\label{fig.5}
\end{figure} \\
Fig.\ref{fig.5} shows the state-of-the-art of the accuracy tests of quantum current sources based on different quantum technologies, along with the best calibration measurement capabilities (CMCs) achieved in national metrology institutes (NMIs) for comparison. It recalls the uncertainties achieved in the present work along with those reported in our previous work \cite{Brun-Picard2016}. This illustrates the wide range of current covered by the quantum current standard. At much lower currents, around 100~pA, uncertainties at the level of $10^{-7}$ are achieved by the best SECS. This uncertainty level is also reached for currents around 1~$\upmu$A by one experiment based on the series connection of quantum Hall resistance array and PJVS \cite{Chae2022}. Let us remark that the uncertainties achieved depend not only on the current source itself but also on the method used to measure the generated current. To this respect, the best known measurement techniques reach relative uncertainties of about $10^{-7}$: \cite{Giblin2019}\cite{SchererULCA2019} (from 100~pA to 1~$\upmu$A), \cite{Chae2020} (around 1~$\upmu$A), \cite{Lee2016} (around 10~mA). On the other hand, the uncertainties $<10^{-8}$ demonstrated with the PQCG comes not only from its own accuracy and stability but also from the measurement with the quantum voltmeter. Providing such an accurate primary quantum current standard in the current range of the best CMCs, which are limited by uncertainties two orders of magnitude larger, is essential both to improve the transfer of the ampere towards end-users and to foster the development of more accurate instruments, as emphasized by the calibration of a digital ammeter with uncertainties limited by the instrument itself. 

Fig.\ref{fig.5} also emphasizes the importance of exploring PQCG capabilities towards even smaller currents, in order to bridge the gap in the current delivered by SECS and devices exhibiting dual Shapiro steps. This would open the way to a new metrological triangle experiment \cite{Likharev1985,Pekola2013}. More precisely, considering the variant of the PQCG proposed in \cite{Poirier2014} and the noise level estimated in this work, we could expect generating and measuring a 10 nA current with a relative uncertainty of $5\times10^{-8}$ after 12~h measurement ($N_1=160$, $I_1=0.33~\upmu$A, $N_2=5400$). \\ \\
\textbf{Future prospects}\\
Another important result is the demonstration of the PQCG accuracy using a 129 times larger resistance (QHRS$_2$) than in \cite{Brun-Picard2016}, which shows its robustness against the load resistance, as required for a true current source. Moreover, the PQCG accuracy being now established, our experiments can be interpreted as calibrations of resistors of 13~k$\Omega$ and 100~$\Omega$ values with a $10^{-8}$ measurement uncertainty. More generally, combining equations (1) and (2) leads to :
\begin{equation}
R_\mathrm{2}=\frac{h}{2e^2}\frac{N_\mathrm{2}}{N_\mathrm{1}}\frac{n_\mathrm{2}}{n_\mathrm{1}}\frac{f_\mathrm{2}^\mathrm{eq}}{f_\mathrm{1}},
\end{equation}
where the equilibrium can be coarsely set by choosing $G$ and $n_{1,2}$ and finely tuned by adjusting $f_1$ and $f_2^{eq}$. This new method combining the PQCG and the quantum voltmeter (see Supplementary Fig.1) paves the way for a paradigm shift for the resistance calibration. It allows to simplify the calibration of a large resistance from $\frac{h}{2e^2}$ to a single step, suppressing the intermediate steps needed using a conventional resistance comparison bridge\cite{Poirier2020}. 

Furthermore, the full quantum instrumentation developed gives foundation to a DC quantum calibrator-multimeter able to provide the primary references of voltage, resistance and current, which are needed in NMIs. This development opens the way to a flexible traceability of electrical measurements to the SI units based on a set of quantum devices gathered in the same experiment. In this perspective, a simplified implementation and operation of the quantum instrumentation is required. The demonstration of the PQCG using a PJVS cooled down in a cryostat equipped with a pulse-tube cryocooler constitutes a first simplification. By gathering the two PJVS and the CCC in the same cryostat, and the two QHRS in a new dry cryostat, our quantum instrumentation will be more practical. Besides, graphene-based single Hall bars \cite{Lafont2015,Ribeiro2015} or arrays \cite{Panna2021,He2022} replacing GaAs devices could provide noise reduction and further simplification owing to an operation under relaxed experimental conditions. In the longer term, QHRS based on the quantum anomalous Hall effect \cite{Fox2018,Gotz2018,Okazaki2022,Patel2024} operating at zero magnetic field like PJVS, or heterostructures-based JJ comprising stacked cuprates~\cite{Martini2024}, could lead to even more compact and easy instrumentation.
\newpage
\Large\textbf{Methods}\\ \\
\large\textbf{Quantum devices}\\ \\
\normalsize
\textbf{Implementation.} The two quantum Hall resistance standards, $\mathrm{QHRS_1}$ and $\mathrm{QHRS_2}$, are both cooled down in the same cryostat at 1.3~K under of magnetic field of 10.8~T. $\mathrm{PJVS_1}$ is cooled down in a small amount of liquid helium maintained at 4.2 K in a recondensing cryostat based on a pulse-tube refrigerator while $\mathrm{PJVS_2}$ is cooled down in a 100 l liquid He Dewar at 4.2 K. The quantization state of PJVS and QHRS is periodically checked following technical guidelines \cite{Rufenacht2018}\cite{Delahaye2003}. In case of occasional trapped flux in the PJVS, a quick heating of the array allows to fully restore the quantized voltage steps. Finally, the CCC is placed in another liquid He Dewar. Connecting five quantum devices, while ensuring quantized operation of PJVS devices and minimizing noise, is very challenging. Many efforts have been spent in optimizing the wiring, the positions of the grounding points and the bias configuration of the PJVS. \\
\textbf{Shielding.} It is essential to cancel the leakage current that could alter the accurate equality of the total currents circulating through the QHRS and the windings. This is achieved by placing high- and low-potential cables (high-insulation $R_L > 1$~T$\Omega$ resistance) connected to the PJVS, to the QHRS and to the CCC inside two separated shields, which are then twisted together and connected to ground. In this way, direct leakage currents short circuiting the QHRS, the most troublesome, are canceled. Other leakage currents are redirected to ground. 
\begin{figure}[ht]
\includegraphics[width=7.5cm]{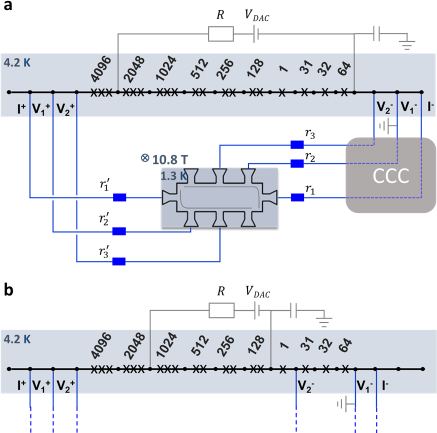}\hfill
\caption{\textbf{Two wiring configurations for PJVS$_1$}. The PJBS is represented by a voltage source $V_{\mathrm{DAC}}$ and a resistance $R=1$ k$\Omega$. The PJBS is connected to ground at the low potential side with of a capacitance $C=1$ $\upmu$F. Black lines : superconducting wires, grey lines : bias lines and connection to ground, blue lines : triple connection. \textbf{a} Configuration with $n_1=4096$. \textbf{b} Configuration with $n_1=1920$ (CCC and QHRS$_1$ are not represented).}\label{fig.PJVS}
\end{figure}\\
\textbf{PJVS cabling.} The two PJVS are binary divided 1 V $\mathrm{Nb/Nb_xSi_{1-x}/Nb}$ series arrays \cite{Behr2012}, both having a total of 8192~JJ and working around 70 GHz. The sequence of the segments that can be biased in the Josephson arrays is the following : 4096/2048/1024/512/256/128/1/31/32/64. Three bondings wires have been added at both ends of the arrays, on the same superconducting pad, in order to implement the triple connection as illustrated in fig.\ref{fig.1}. A 50~$\Omega$ heater is placed close to the Josephson array chip allowing to get rid of trapped flux within few minutes. We have used a prototype of the hermetic cryoprobe developed for the recondensing cryostat, which had only 8 available wires, reducing the possible wiring configurations for $\mathrm{PJVS_1}$. Fig.\ref{fig.PJVS}a and b show the two wiring configurations used with $n_1=4096$ and $n_1=1920$, respectively. They differ essentially on the way the triple connection is implemented at the low potential side of the bias source (PJBS). In Fig.\ref{fig.PJVS}a, the triple connection is done on the same superconducting pad, while in Fig.\ref{fig.PJVS}b, JJ are present between the bias wire and the wires connecting $\mathrm{QHRS_1}$, but also between the second and third connection of the triple connection. If the current circulating in the JJ is less than half the amplitude of the $n=0$ Shapiro step ($< 500$~$\upmu$A), both configurations are equivalent. However, the second one turned out to be less reliable than the first one when connecting the rest of the circuit. Because of the ground loop including the JJ, it was very sensitive to trapping magnetic flux. Nonetheless, the results of Fig.\ref{DataCompl1et2}c show that quantized currents can be generated using this configuration. It was used for the calibration of the DA. The best accuracy tests reported in Fig.\ref{fig.2} and Fig.\ref{DataCompl1et2}a were done with the first configuration. \\ \\
\large\textbf{CCC for correction-free PQCG}\\
\normalsize
\begin{figure}[ht]
\includegraphics[width=8.0cm]{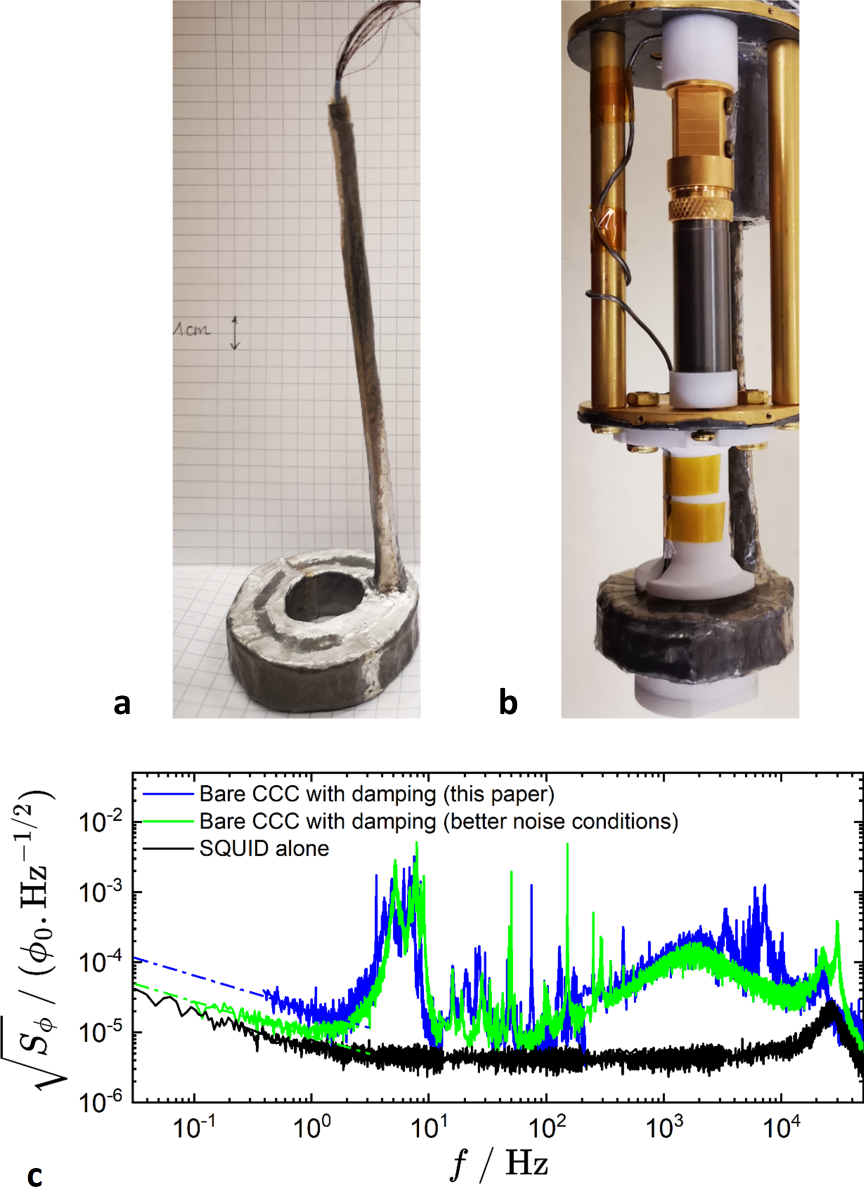}\hfill
\caption{\textbf{CCC details}. \textbf{a} Photo of the CCC. \textbf{b} Photo of the CCC mounted in the cryogenic probe. (c) Flux noise measured at the output of the SQUID alone (black line) and coupled to the bare CCC with damping circuit in this work (blue line) and in best conditions (green line). Blue and green dash dotted lines present $1/\sqrt{f}$ slopes.}\label{fig.CCC}
\end{figure}\\
\textbf{Design.} The new CCC (Fig.\ref{fig.CCC}a and b) is made of 20 windings of 1, 1, 1, 2, 2, 16, 16, 16, 32, 64, 128, 128, 160, 160, 465, 465, 1600, 1600, 2065 and 2065 turns, with a total number of 8789 turns. They are embedded in a superconducting toroidal shield made of 150 $\upmu$m thick Pb foils, forming three electrically isolated turns to prevent non-ideal behaviour at the ends of the shield. The architecture is inspired from the design of a CCC used in a quantum Hall resistance bridge \cite{Poirier2020} (enabling ratios close to 1.29), but with 5 additional windings. The triple connection is possible for the windings of 1, 2, 16, 128, 160, 465 and 1600 turns. The dimensions have been chosen to be mounted on a cryogenic probe designed to be compatible with a 70~mm diameter neck of a liquid He Dewar. The inner and outer diameter of the toroidal shield are 19~mm and 47~mm, respectively. The chimney is about 125~mm high. The CCC is enclosed in two successive 0.5 mm thick Pb superconducting cylindrical screens and in a Cryoperm shield surrounding the whole, corresponding to an expected overall magnetic attenuation of about 200~dB. It is equipped with a Quantum Design Inc. DC SQUID, placed in a separate superconducting Nb shield, and coupled to the CCC via a superconducting flux transformer composed of a wire wound sensing coil placed as close as possible to the inner surface of the CCC. The CCC sensitivity $\gamma_\mathrm{CCC}=8$ $\upmu$A$\cdot$turn/$\upphi_0$ has been maximized with a sensing coil of 9 turns compatible with the geometrical constraints. The 20 windings are connected by 40 copper alloy wires (AWG 34) placed in a stainless steel shield.\\
\textbf{Noise.} Fig.\ref{fig.CCC}c shows the noise spectrum at the output of the SQUID. The base noise level of the CCC (green line) amounts to 10 $\upmu\upphi_0/\sqrt{\mathrm{Hz}}$ at 1 Hz in the best noise conditions, slightly higher than the white noise level of 3 $\upmu\upphi_0/\sqrt{\mathrm{Hz}}$ of the SQUID alone (black line). Below 1 Hz, one can observe an increased noise compatible with a 1/$f$ noise contribution. Resonance peaks present at frequencies below 10~Hz are certainly due to an imperfect decoupling of the Dewar from the ground vibrations \cite{Poirier2020}.\\
\textbf{Damping circuit.} A damped resonance at 1.6~kHz is due to the use of a damping circuit to improve the stability of the feedback loop. The damping circuit is made of a $C_\mathrm{D}=100$~nF capacitance at room temperature in series with a $R_\mathrm{D}=$ 1~k$\Omega$ resistor and a $N_\mathrm{D}=$ 2065 turns CCC winding at $T_\mathrm{D}=$ 4.2~K. It strongly damps the CCC resonances (around 10 kHz) which are excited by the external noise captured. The counterpart is an increase of the magnetic flux noise detected by the SQUID around 1.6~kHz caused by the Johnson-Nyquist noise emitted by the resistor. However, placing the resistor at low temperature reduced the noise magnitude by a factor of ten compared to the previous experiment \cite{Brun-Picard2016}, with a maximum flux noise of $\frac{N_\mathrm{D}}{\gamma_\mathrm{CCC}}\sqrt{\frac{4k_\mathrm{B}T_\mathrm{D}}{R_\mathrm{D}}}\simeq$ 124~$\upmu\upphi_0/\sqrt{\mathrm{Hz}}$ at 1.6~kHz \cite{Djordjevic2021}. In the experimental conditions of this paper, the noise was measured slightly higher, as described by blue curve in Fig.\ref{fig.CCC}c. The noise level at 1 Hz rises to about 20 $\mathrm{\upmu\upphi_0/Hz^{1/2}}$. \\
\textbf{Current sensitivity.} It is related to the flux generated by the screening current circulating on the shield and can be estimated from the CCC noise spectrum and $\gamma_\mathrm{CCC}$, it corresponds to 80 pA$\cdot$ turns/$\sqrt{\mathrm{Hz}}$ at 1 Hz in the best conditions. \\
\textbf{Accuracy.} The CCC accuracy can be altered by magnetic flux leakage detected by the pickup coil. It can be tested by series opposition measurements of windings of identical number of turns. In the best noise conditions, using a measurement current of 30 to 100~mA, the ratio error on the number of turns $\delta N/N$ was measured $<10^{-9}$ for $N$ considered in this paper. However, slightly larger errors were measured at the time of this work leading to a CCC contribution to the PQCG Type B uncertainty of $2\times10^{-9}$ (see Type B uncertainty section). \\\\
\large\textbf{Multiple series connection}\\ \\
\normalsize
In the multiple series connection (see fig.\ref{fig.1} and fig.\ref{fig.PJVS}a), the series resistances of the connections result in an effective resistance, which adds to the quantized Hall resistance \cite{Delahaye1993,Poirier2014}. This leads to a lower value of the quantized current $I_\mathrm{PQCG}=Gn_{\mathrm 1}ef_{\mathrm 1}(1-\alpha_n)$, where $\alpha_n$ is positive and exponentially decreasing with the number of connections $n$. The series resistances $r_{1}$, $r_{2}$, $r_{3}$, $r^{'}_{1}$, $r^{'}_{2}$ and $r^{'}_{3}$, as indicated in fig.\ref{fig.PJVS}a, one calculates, using a Ricketts and Kemeny model \cite{Ricketts1988} of the Hall bar, $\alpha_2=\frac{r_{1}r_{2}}{R_\mathrm{H}^2}+\frac{r^{'}_{1}r^{'}_{2}}{R_\mathrm{H}^2}$ and $\alpha_3=\frac{r_{1}r_{2}r_{3}}{R_\mathrm{H}^3}+\frac{r^{'}_{1}r^{'}_{2}r^{'}_{3}}{R_\mathrm{H}^3}$ for the double series connection and the triple series connection, respectively. The series resistances to be considered are those of the QHRS contacts and cables, those of the long cables linking the quantum devices, and those of the different combinations of CCC windings necessary to obtain the desired number of turns $N_1$. For all the experiments based on the triple connection technique, $\alpha_3$ is calculated below $1.5\times10^{-10}$, except for the measurement performed with a 50 $\Omega$ resistor added in series with the first CCC winding (see fig.\ref{fig.2}e and fig.\ref{DataCompl1et2}b), which results in $\alpha_3=(1.44\pm 0.023)\times10^{-9}$. For the measurement reported in fig.\ref{fig.2}e and in fig.\ref{DataCompl1et2}b using the double connection scheme, one calculates $\alpha_2=(2.344\pm 0.037)\times10^{-7}$.\\
\large\textbf{Uncertainties, weighted mean values, combined results and errors bars}\\ \\
\normalsize
Two types of measurement uncertainties are considered: the Type A uncertainties which are evaluated by statistical methods, the Type B uncertainties evaluated by others methods. \\ \\
\textbf{Type A uncertainty.} The type A uncertainty for one measurement of $\Delta I/I$, $u^\mathrm{A}$, is given by : 
\begin{equation}
u^\mathrm{A}=\frac{u^\mathrm{A}(f_\mathrm{2}^\mathrm{eq})}{f_\mathrm{1}}
\end{equation}
where 
\begin{equation}
u^\mathrm{A}(f_\mathrm{2}^\mathrm{eq})=\frac{(f_\mathrm{2}^+-f_\mathrm{2}^-)\sqrt{(w^\mathrm{A}_{+}\overline{\Delta V_{f_2^{-}}})^2+(w^\mathrm{A}_{-}\overline{\Delta V_{f_2^{+}}})^2}}{(\overline{\Delta V_{f_2^{+}}}-\overline{\Delta V_{f_2^{-}}})^2}
\end{equation}
with $w^\mathrm{A}_{+}$ and $w^\mathrm{A}_{-}$ the standard uncertainties (coverage factor $k=1$) of the mean of voltage series $\Delta V_{f_2^{+}}$ and $\Delta V_{f_2^{-}}$, respectively. Their calculation is legitimated by the time dependence of the Allan deviation which demonstrates a dominant white noise (Supplementary Fig.2b).

To account for the intra-day noise observed in measurements reported in Fig.\ref{fig.2}e and g, a Type A uncertainty, $u^\mathrm{A}_\mathrm{id-noise}=10^{-8}$, is added to each data points. Its value was determined so that the $\chi^2=\frac{1}{N}\Sigma_1^N \frac{(X_i-\overline{X})^2}{u(X_i)^2}<1$ criterion is fulfilled, where here $N$ is the number of values, $X_i$ the value, $\overline{X}$ the weighted mean of all $X_i$ and $u(X_i)$ the standard uncertainty of the $X_i$. As suggested by an investigation of the quality of the power line of our laboratory, the observed intra-day noise could be caused by a recent increase of the electrical noise pollution in the main line (back-up power supply) supplying the laboratory. We have recently revealed an increase of the voltage noise between the ground and the neutral, probably caused by an increase power consumption by non-linear equipments. This random noise increase could lead to extra noise in SQUID measurements. This investigation is in course. \\ \\
\textbf{Type B uncertainty.} Table \ref{Type-B PQCG} shows the basic Type B standard uncertainty budget of the accuracy test, which includes contributions of both the PQCG and the quantum voltmeter. The implementation of the triple series connection and the use of new measurement protocols based on the adjustment of the current using only the Josephson parameters have cancelled (current divider) or strongly reduced (cable correction) the most important contributions of the previous experiment \cite{Brun-Picard2016}. We would expect a total uncertainty below $1\times 10^{-9}$. However, it turns out that we measured, momentarily during the measurement campaign, CCC ratio errors slightly higher than the typical values. The ratio errors ranged from $2\times 10^{-9}$ to $0.1\times 10^{-9}$ for windings of number of turns from 128 to 1600, respectively, which were used either for $N_1$ or $N_2$ in the experiment reported here. Using a conservative approach, we have therefore considered here a Type B uncertainty of $2\times10^{-9}$ for the CCC, which dominates the uncertainty budget. Other components were detailed in \cite{Brun-Picard2016}. The SQUID electronic feedback is based on the same SQUID type and same pre-amplifier from Quantum Design. It is set in the same way using a $4.2~V/\upphi_0$ close-loop gain whatever the number of turns $N_2$ used. The VCCS current source preadjusts the output current, such that the SQUID feedbacks only on a small fraction lower than $2\times 10^{-5}\times I_\mathrm{PQCG}$ (see \cite{Brun-Picard2016}). Owing to the cable shields, leakage to ground are redirected to ground, i.e. parallel to CCC winding \cite{Poirier2014}. The current leakage error amount to $r_1/R_{L}$, which is below $7\times10^{-12}$ in our experiments. It results a relative Type B uncertainty of $2\times 10^{-9}$ for the PQCG and a relative Type B uncertainty of $u_\mathrm{B}=2.1\times 10^{-9}$ for the accuracy test.
\begin{table}[ht]
\caption{\textbf{Basic relative Type B standard uncertainty budget of the accuracy test.} It includes contributions of the PQCG and of the quantum voltmeter. Bold values correspond to total contributions (PQCG, quantum voltmeter, accuracy test). The coverage factor is $k=1$.}\label{Type-B PQCG}
\begin{center}
\begin{tabular}{|c|c|}
  \hline
 \textbf{Contribution}&$\bm{u^\mathrm{B}}$\\
                      &$\bm{(10^{-9})}$\\ \hline
  Triple series connection &0.2\\
  Electronic feedback &$<0.5$ \\
  CCC accuracy &$<2$\\
  QHRS$_1$ &$<0.1$ \\
  PJVS$_1$ &$<0.1$ \\
  Current leakage &$<0.01$ \\
  Frequency &$<0.01$\\ \hline
  \textbf{PQCG} &\textbf{2}\\ \hline
  QHRS$_2$ &$<0.1$ \\
  PJVS$_2$ &$<0.1$ \\
  Null detector &$<0.5$ \\ \hline
  \textbf{Quantum voltmeter} &\textbf{0.5} \\ \hline
  \textbf{Accuracy test}  & \textbf{2.1} \\ \hline
\end{tabular}

\end{center}
\end{table} \\ \\
\textbf{Weighted mean values}.
Weighted mean values, $\Delta I/I_\mathrm{WM}$ and their uncertainties, $u^\mathrm{A}_\mathrm{WM}$, are calculated from the $\Delta I/I$ series values and their Type A uncertainties.\\
In fig.\ref{fig.2}b and d, they are given by:
\begin{align}
\Delta I/I_\mathrm{WM}&=\frac{\sum_j \Delta I/I_j\times\frac{1}{(u^\mathrm{A}_j)^2}}{\sum_j\frac{1}{(u^\mathrm{A}_j)^2}}\\
u^\mathrm{A}_\mathrm{WM}&=\frac{1}{\sqrt{\sum_j\frac{1}{(u^\mathrm{A}_j)^2}}}
\end{align}\\
In fig.\ref{fig.2}f and h, they are given by:
\begin{align}
\Delta I/I_\mathrm{WM}&=\frac{\sum_j \Delta I/I_j\times\frac{1}{(u^\mathrm{A}_j)^2+(u^\mathrm{A}_\mathrm{id-noise})^2}}{\sum_j\frac{1}{(u^\mathrm{A}_j)^2+(u^\mathrm{A}_\mathrm{id-noise})^2}}\\
u^\mathrm{A}_\mathrm{WM}&=\frac{1}{\sqrt{\sum_j\frac{1}{(u^\mathrm{A}_j)^2+(u^\mathrm{A}_\mathrm{id-noise})^2}}}
\end{align}
where $u^\mathrm{A}_\mathrm{id-noise}=10^{-8}$ is the additional Type A component added to each data point to take account of the intra-day noise.\\ \\ \\
\textbf{Combined results $\overline{\Delta I/I}$}.\\
For accuracy tests performed at 11.48 $\upmu$A using $N_1=465$ and at 45.94 $\upmu$A using $N_1=465$, the combined result $\overline{\Delta I/I}$ is the mean value of the $\Delta I/I_\mathrm{WM}$ values obtained using the measurement protocol I$_+$ and I$_-$: $\overline{\Delta I/I}=(\Delta I/I_\mathrm{WM}(\mathrm{I}_+) + \Delta I/I_\mathrm{WM}(\mathrm{I}_-))/2$.\\
For accuracy tests performed at 5.74 $\upmu$A using $N_1=160$, at 45.94 $\upmu$A using $N_1=160$, at 57.42 $\upmu$A using $N_1=160$ and at 43.07 using $n_1=1920$, the combined result $\overline{\Delta I/I}$ is the weighted mean value calculated from the values $(\Delta I/I_\mathrm{WM}(\mathrm{I}_+) + \Delta I/I_\mathrm{WM}(\mathrm{I}_-))/2$ and $\Delta I/I_\mathrm{WM}(\mathrm{I}_\pm)$ and their respective Type A uncertainties.\\ \\  
\textbf{Combined uncertainties}. The combined uncertainty, $u^\mathrm{c}_\mathrm{WM}$, is given by $u^\mathrm{c}_\mathrm{WM}=\sqrt{(u^\mathrm{A}_\mathrm{WM})^2+(u^\mathrm{B})^2}$. The combined uncertainty of $\overline{\Delta I/I}$ is given by $u^\mathrm{c}=\sqrt{(u^\mathrm{A}_{\overline{\Delta I/I}})^2+(u^\mathrm{B})^2}$\\ \\
\textbf{Error bars}. Error bars in the different figures represent measurement uncertainties corresponding to one standard deviation (i.e. $k=1$). This means an interval of confidence of 68\% if a gaussian distribution law is assumed. These measurement uncertainties are either Type A uncertainties or combined uncertainties.\\
\underline{Figure~\ref{fig.2}}\\
In a, c, e and g, error bars correspond to only Type A uncertainties, $u^\mathrm{A}$.\\
In b, d, f and g, errors bars correspond to combined uncertainties, $u^\mathrm{c}_\mathrm{WM}$\\
\underline{Figure \ref{DataCompl1et2}}\\
In a, error bars correspond to combined standard uncertainties $\sqrt{(u^\mathrm{A})^2+(u^\mathrm{B})^2}$.\\
In b, error bars correspond to the combination of Type A standard uncertainties according to $\sqrt{(u^\mathrm{A})^2+u^\mathrm{A}(\Delta I_\mathrm{ref}/I_\mathrm{ref})^2}$. \\
In c, error bars correspond to $u^\mathrm{A}$.\\
\underline{Figure~\ref{fig.3}}\\
In a and b, errors bars correspond to uncertainties $u(u^\mathrm{A}_\mathrm{exp})$, which are standard deviations of $u^\mathrm{A}$ values, and not standard deviations of the means, in order to reflect the ranges over which the noise levels vary. \\
\underline{Figure~\ref{fig.4}}\\
In a and b, error bars correspond to Type A standard uncertainties.\\
\underline{Figure~\ref{fig.5}}\\
In inset of Fig.\ref{fig.5}a, error bars correspond to combined standard uncertainties, $u^\mathrm{c}$. \\ \\
\large\textbf{Experimental settings of the PQCG used during the accuracy tests}\\ \\
\normalsize
All accuracy tests are performed using the PQCG settings reported in Table \ref{SettingAccuracyTest}, except one measurement at 5.74~$\upmu$A using frequencies $f_1=f_2=70.02$~GHz (see Fig.2a) and one measurement at 45.94 $\upmu$A which uses different frequencies $f_1=70$ GHz and $f_2=69.999976$~GHz, respectively, to accommodate for the deviation of a few parts in $10^7$ of the PQCG current from equation (1) caused by the implementation of the double connection only (see Fig.\ref{fig.2}e). Measurements are carried out with $\Delta f$=40 or 80 kHz.
\begin{table}[ht]
\caption{\textbf{Experimental settings of the PQCG for the accuracy tests.}\label{SettingAccuracyTest}}
\scriptsize
\begin{center}
\begin{tabular}{|c|c|c|c|c|c|c|c|}
 \hline
  Current&$I_\mathrm{PQCG}$&$n_1$&$f_1=f_2$&$I_1$&$G$=$N_1$/$N_2$&$n_2$&$V_2$\\
  value&($\upmu$A)& &(GHz)&($\upmu$A)& &  &(V)\\ \hline
  @ 5.74&5.742201056&4096&70&45.94&160/1280&512&0.074\\
  @11.48&11.48440211&4096&70&45.94&465/1860&1024&0.148\\
  @43.07&43.06650792&1920&70&21.53&160/80&3840&0.556\\
  @45.94&45.93760845&4096&70&45.94&160/160&4096&0.593\\
  @45.94&45.93760845&4096&70&45.94&465/465&4096&0.593\\ 
  @57.42&57.42201056&4096&70&45.94&160/128&5120&0.741\\ \hline
\end{tabular}
\end{center}
\end{table}\\ \\
\large\textbf{Experimental, $u^\mathrm{A}_{\mathrm{exp}}$ and calculated, $u^\mathrm{A}_{\mathrm{calc}}$, standard uncertainties.}\\ \\
\normalsize
The uncertainties $u^\mathrm{A}_{\mathrm{exp}}(\tau_m)$ are calculated by averaging the uncertainty values, $u^\mathrm{A}$, of each series, after normalization to the same measurement time $\tau_m=N_s\tau_0$ where $N_s=16$ is the number of sequences, and to the same measurement protocol I$_\pm$. The standard deviation, $u(u^\mathrm{A}_\mathrm{exp})$, is calculated from the different values of a series. $u^\mathrm{A}_{\mathrm{calc}}(\tau_m)$ is calculated using the relationship $u^\mathrm{A}_\mathrm{calc}(\tau_m=N_s\tau_0)=\frac{1}{\sqrt{N_s}}\sqrt{\frac{3}{8}}\sqrt{\frac{S_{\Delta I/I}}{2\tau_A}}$, where $\tau_A=12$~s is the acquisition time for one single voltage measurement (Supplementary Fig.2a) and $\sqrt{S_{\Delta I/I}}$ is the noise density of $\Delta I/I$. More precisely, $\sqrt{\frac{S_{\Delta I/I}}{2\tau_A}}$ is the relative standard deviation corresponding to the acquisition time $\tau_A$ assuming an effective white noise density. The pre-factor $\sqrt{\frac{3}{8}}$ comes from the combination of the standard deviations corresponding to the measurement protocol I$_\pm$, where the voltage of each sequence is obtained by $[(\Delta V (+I)_1+\Delta V (+I)_3)/2-\Delta V (-I)_2]$, with $\Delta V (+I)_1$, $\Delta V (-I)_2$ and $\Delta V (+I)_3$ three successive voltage acquisitions, performed with positive current, then negative current, and then positive current (Supplementary Fig.2a). Finally, the factor $\frac{1}{\sqrt{N_s}}$ comes from the white noise hypothesis justified by Supplementary Fig.2b. The noise density $\sqrt{S_{\Delta I/I}}$ is given by:
\begin{equation}
\sqrt{S_{\Delta I/I}}=\sqrt{\frac{S_V}{V_2^2}+(\frac{1}{N_1I_1}\gamma_\mathrm{CCC})^2S_{\upphi}+\frac{4k_\mathrm{B}T}{R_1I_1^2}},
\end{equation}
where $k_\mathrm{B}$ is the Boltzmann constant and $T$=1.3 K. Three noise contributions are considered: the voltage noise power spectral density, $S_V$, of the quantum voltmeter, which includes the noise of the null detector and some external voltage noise captured, the magnetic flux noise power spectral density, $S_{\upphi}$, detected by the SQUID, which includes the SQUID noise and some external magnetic flux noise captured and the Johnson-Nyquist noise power spectral density emitted by the resistor $R_1$ in the primary loop. The Johnson-Nyquist noise of the resistor $R_2$ is included in $S_V$. The third term contributes to $\sqrt{S_{\Delta I/I}}$ by about $1.6\times10^{-9}/\mathrm{Hz}^{1/2}$ for $I_1=45.94$ $\upmu$A, leading to a negligible uncertainty contribution of $5\times10^{-11}$ for a measurement time $\tau_m=16\tau_0$. This third term is therefore not considered in our calculations. Values reported in Fig.\ref{fig.2}e and f are calculated using a voltage noise of $\sqrt{S_V}=\mathrm{28~nV/Hz^{1/2}}$ and a magnetic flux noise detected by the SQUID of $\sqrt{S_{\upphi}}=6.2\times10^{-5}\upphi_0/\mathrm{Hz}^{1/2}$. \\ \\ 
\large\textbf{Measurement protocol for the ammeter calibration}\\ \\
\normalsize
Calibrations of the ammeter HP3458A are performed using settings of the PQCG reported in Supplementary Table.1. In these experiments, output current values are changed by varying both the gain $G$ and the frequency $f_1$. Using the PQCG to perform ammeter calibration consists in replacing $\mathrm{QHRS_2}$ by the device under test and removing the quantum voltmeter. The connection is done through a low-pass filter (highly insulated PTFE 100 nF on the differential input). A common mode torus has also been introduced to minimize the noise. Supplementary Fig.3a shows recordings by the ammeter (HP3458A) as a function of time for several alternations of $I_\mathrm{PQCG}$ at 107.666272 $\upmu$A using the measurement protocol I$_+$. The acquisition time, the waiting time are of 10 s and 2 s, respectively. The measured current is determined from the average of the values obtained for several measurement groups. The time dependence of the Allan deviation, reported in Supplementary Fig.3b, shows that the standard deviation of the mean is a relevant estimate of the Type A relative uncertainty at $\tau_m=144$ s. An uncertainty of $2\times10^{-7}$ is typically achieved after a total measurement time of 144 s.\\ \\
\Large{\textbf{Data Availability}} \\ 
\normalsize
Source data for figures 2, 3, 4, 5 and 6, which support the findings of this study, are provided with the paper.\\
\providecommand{\noopsort}[1]{}\providecommand{\singleletter}[1]{#1}%
 
\newpage
\Large{\textbf{Acknowledgments}} \\ 
\normalsize{We wish to acknowledge Mohammed Mghalfi for its technical support. We thank Daniel Est\`eve (CEA/SPEC, France), Christian Glattli (CEA/SPEC, France), Yannick De Wilde (ESPCI, France) and Almazbek Imanaliev (LNE, France) for their critical reading and comments. Part of this work was supported by the project 23FUN05 AQuanTEC. The project 23FUN05 AQuanTEC has received funding from the European Partnership on Metrology, co-financed from the European Union’s Horizon Europe Research and Innovation Programme and by the Participating States.}\\ \\
\Large{\textbf{Author contributions}}\\
\normalsize{S. D. and W. P. planned the experiments. R. B performed the cabling and the characterizations of the programmable Josephson standards fabricated by PTB. S. D. and W. P. developed the instrumentation, conducted the electrical metrological measurements, analyzed the data and wrote the paper. All authors contributed to the final version.}\\ \\
\Large{\textbf{Competiting interests}}\\
\normalsize{The authors declare no competiting interest.}\\ \\
\newpage
\onecolumngrid
\textbf{\Large{Supplementary Information}}\\ \\
\normalsize
\renewcommand{\figurename}{\textbf{Supplementary Fig.}}
\renewcommand{\thefigure}{\arabic{figure}}
\renewcommand{\tablename}{\textbf{Supplementary Table}}
\renewcommand{\thetable}{\arabic{table}}
\setcounter{figure}{0}
\setcounter{table}{0}
\begin{figure}[h]
\includegraphics[width=11.0cm]{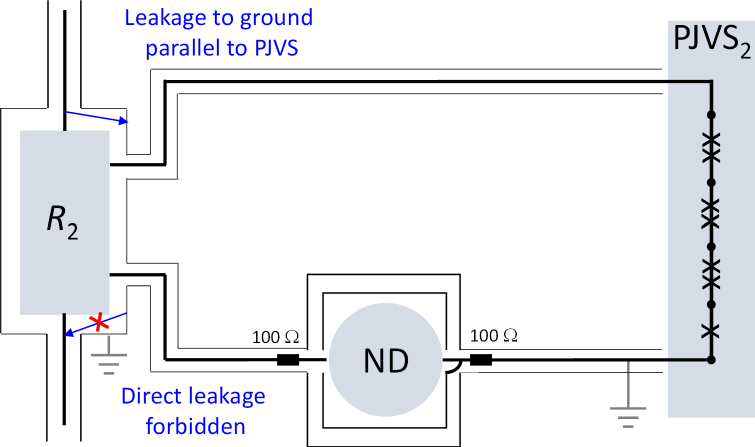}\hfill
\caption{\textbf{Quantum voltmeter}. Scheme of the quantum voltmeter measuring the voltage drop at the terminal of a resistor of resistance $R_2$. It is made of a second PJVS, PJVS$_2$ and of nanovoltmeter (EM Electronics Model N11). Thin lines represent the outer shielding of the cables, which is connected to ground. The low potential of PJVS$_2$ is also connected to ground. At equilibrium, direct current leakage parallel to $R_2$ is strongly screened by the shielding: no current can circulate between the low potential of $R_2$ and ground because both are at the same potential. All current leakage to ground is deviated in parallel to PJVS$_2$. One 100 $\Omega$ resistance is added in series on each side of the ND to limit the circulation of strong current during current reversal.}\label{FigQV}
\end{figure}
\newpage
\begin{figure}[h]
\includegraphics[width=11.0cm]{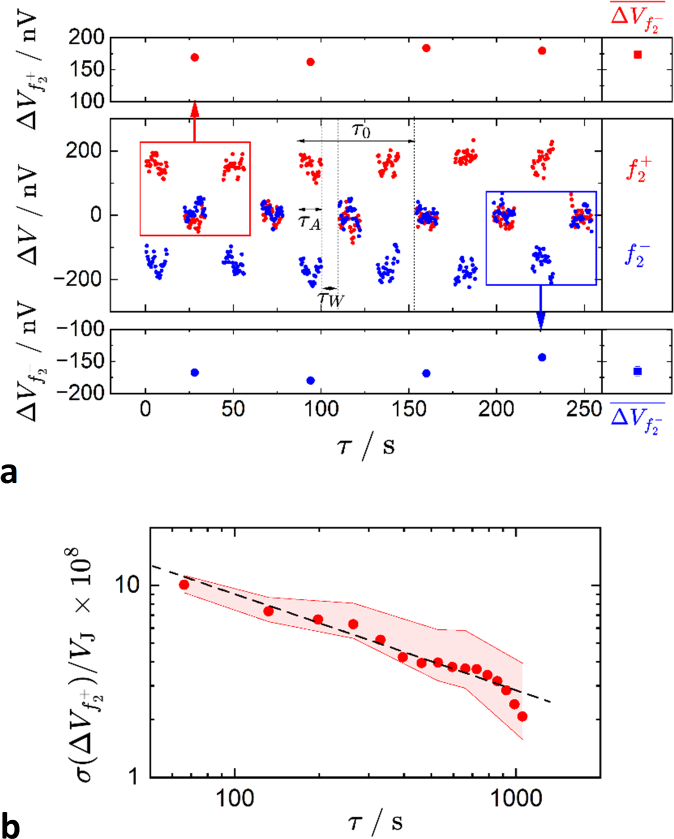}\hfill
\caption{\textbf{Measurement protocol for the accuracy test of the PQCG}. \textbf{a} ND readings, $\Delta V$, as a function of time, $\tau$, obtained using the measurement protocol I$_+$ at $11.48440211$ $\upmu$A, during the two steps carried out at $f_2^{+}$ (red circle) and $f_2^{-}$ (blue circle), respectively. Here, $f_1=f_2=70$ GHz and $\Delta f=80$ kHz. At each alternation of $I_\mathrm{PQCG}$, a waiting time $\tau_W$ of 10 s is applied, and the acquisition time in each state is $\tau_A$=12 s. The period time is $\tau_0=66$ s. Mean values, $\overline{\Delta V_{f_2^{-}}}$ and $\overline{\Delta V_{f_2^{+}}}$ are then obtained by averaging the series of voltage values $\Delta V_{f_2^{-}}$ and $\Delta V_{f_2^{+}}$. Each of these voltage values is itself given by $[(\Delta V (+I)_1+\Delta V (+I)_3)/2-\Delta V (0)_2]$, where $\Delta V (+I)_1$, $\Delta V (0)_2$ and $\Delta V (+I)_3$ are three successive voltage measurements performed with positive current switched on, then off, and then on (rectangle). \textbf{b} Relative Allan deviation (red points) calculated from a series of 32 values of $\Delta V_{f_2^{+}}$ plotted as a function of $\tau$. The light-red area corresponds to the 68.3 \% confidence interval. The good agreement of the $9\times10^{-7}\times\tau^{-1/2}$ fit (black dashed line) with data confirms the dominant contribution of the white noise. This legitimate the calculation of standard deviation of the mean to estimate the uncertainties.}\label{FigProto}
\end{figure}
\newpage
\begin{table}[h]
\caption{\textbf{Experimental settings of the PQCG for the ammeter calibration.\label{SettingAmmeter}}}
\begin{center}
\begin{tabular}{|c|c|c|c|c|c|}
  \hline
 Current&$I_\mathrm{PQCG}$&$n_1$&$f_1$&$I_1$&$G$=$N_1$/$N_2$\\
 value&($\upmu$A)& &(GHz)&($\upmu$A)& \\ \hline
  @ 62.58&062.581019&1920&70&21.53&465/160\\
  @107.64&107.635508&1920&69.98&21.53&465/93\\
  @107.65&107.653965&1920&69.992&21.53&465/93\\
  @107.67&107.666270&1920&70&21.53&465/93\\
  @107.68&107.684727&1920&70.012&21.54&465/93\\
  @107.70&107.697032&1920&70.02&21.54&465/93\\
  @107.67&107.666270&1920&70&21.53&160/32\\ \hline
\end{tabular}
\end{center}
\end{table}
\newpage
\begin{figure}[h]
\includegraphics[width=11cm]{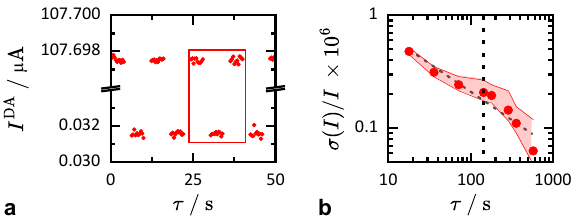}\hfill
\caption{\textbf{Measurement protocol for the ammeter calibration.} \textbf{a} Data recorded by the HP3458A of $I_\mathrm{PQCG}=107.666272$ $\upmu$A as a function of time, $\tau$, for measurement protocol I$_+$. \textbf{b} Time dependence of the Allan deviation. The vertical dash-line is at $\tau=144$ s.}\label{TraceHPAllan}
\end{figure}
\end{document}